# Realizing polarization conversion and unidirectional transmission by using a uniaxial crystal plate

Xiaohu Wu and Ceji Fu[a]

*LTCS and Department of Mechanics and Engineering Science, College of Engineering, Peking University, Beijing 100871, China*

We show that polarization states of electromagnetic waves can be manipulated easily using a single thin uniaxial crystal plate. By performing a rotational transformation of the coordinates and controlling the thickness of the plate, we can achieve a complete polarization conversion between TE wave and TM wave in a spectral band. We show that the off-diagonal element of the permittivity is the key for polarization conversion. Our analysis can explain clearly the results found in experiments with metamaterials. Finally, we propose a simple device to realize unidirectional transmission based on polarization conversion and excitation of surface plasmon polaritons.

[a] Corresponding author, Email: cjfu@pku.edu.cn

Polarization is one of the important characteristics of electromagnetic (EM) waves.[1] In the past ten years, how to manipulate the polarization of EM waves has become a hot research frontier and various schemes have been proposed. Hao et al.[1] realized complete polarization conversion through wave reflection by an anisotropic metamaterial (MM) plate in 2007. After that, investigations on manipulation of the polarization of EM waves have been conducted extensively using MMs.[2-17] However, these MMs usually demand complex nanostructures, which are difficult to fabricate for working in the optical region. The method of transformation optics (TO), which can result in a transformation on the material's permittivity and permeability, was applied by Pendry[18] and Leonhardt[19] to realize cloaking in 2006. It has now become an important method for designing different types of nonreciprocal and switching devices.[20-22] In 2013, Liu et al.[23] combined TO and field transformation (FT) to realize complete polarization conversion between transverse electric (TE) and transverse magnetic (TM) waves. Unfortunately, this method also needs MMs to fulfill the required flexible optical properties. Therefore, it is an urgent necessity to find an easy way to realize complete conversion of polarization between TE and TM waves. Here we propose an alternative approach to manipulate the polarization of EM waves by using a uniaxial crystal plate based on TO. Through adjusting the orientation of the optical axis and changing the thickness of the plate, we can achieve complete conversion between TE wave and TM wave within a certain band. This method provides a straight and clear way for understanding the essence of polarization conversion in the uniaxial crystal, as well as in MMs[1,3-5] previously studied. On the other hand, polarization conversion has been successfully used to achieve unidirectional transmission with bianisotropic metasurfaces[24], MMs[25-28], magneto-optical materials[29] and chiral structures[30]. Based on the polarization conversion in uniaxial crystal, we also propose a simple device to realize unidirectional transmission and the difference of transmission in opposite directions can reach 0.78. In our case, neither the MMs nor the magneto-optical materials are needed.

We take hexagonal boron nitride (hBN) as an example in this work. By setting the y-axis of the coordinates parallel to the optical axis of hBN, the permittivity tensor of the uniaxial crystal can be described by the Lorentz model as[31]:

$$\overline{\overline{\varepsilon}} = \begin{pmatrix} \varepsilon_\perp & 0 & 0 \\ 0 & \varepsilon_\parallel & 0 \\ 0 & 0 & \varepsilon_\perp \end{pmatrix}, \quad \varepsilon_m = \varepsilon_{\infty,m}\left(1 + \frac{\omega_{LO,m}^2 - \omega_{TO,m}^2}{\omega_{TO,m}^2 - \omega^2 + j\omega\Gamma_m}\right) \quad (1)$$

where $m = \perp, \parallel$ indicates the component perpendicular or parallel to the optical axis. $\omega$ is the wavenumber, the other parameters are $\omega_{TO,\perp} = 1370$ cm$^{-1}$, $\omega_{TO,\parallel} = 780$ cm$^{-1}$, $\omega_{LO,\perp} = 1610$ cm$^{-1}$, $\omega_{LO,\parallel} = 830$ cm$^{-1}$, $\varepsilon_{\infty,\perp} = 4.87$, $\varepsilon_{\infty,\parallel} = 2.95$, $\Gamma_\perp = 5$ cm$^{-1}$ and $\Gamma_\parallel = 4$ cm$^{-1}$.

After performing a rotational transformation of the coordinates around the z-axis by an angle $\alpha$ (see Fig.1), we get a new permittivity tensor

$$\overline{\overline{\varepsilon}} = \begin{pmatrix} \varepsilon_{xx} & \varepsilon_{xy} & 0 \\ \varepsilon_{yx} & \varepsilon_{yy} & 0 \\ 0 & 0 & \varepsilon_{zz} \end{pmatrix} \quad (2)$$

where $\varepsilon_{xx} = \varepsilon_\parallel + (\varepsilon_\perp - \varepsilon_\parallel)\cos^2\alpha$, $\varepsilon_{xy} = \varepsilon_{yx} = 0.5(\varepsilon_\perp - \varepsilon_\parallel)\sin 2\alpha$, $\varepsilon_{yy} = \varepsilon_\parallel + (\varepsilon_\perp - \varepsilon_\parallel)\sin^2\alpha$ and $\varepsilon_{zz} = \varepsilon_\perp$. Comparing with the old permittivity tensor

(1), the new permittivity tensor gains its off-diagonal element $\varepsilon_{xy}$, which is the key for polarization conversion between TE and TM waves, as shown below.

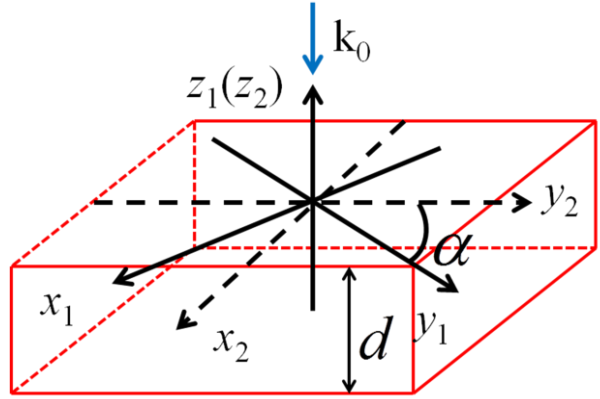

Fig.1 (Color online) Skematic of the coordinates and the uniaxial crystal plate of thickness $d$. Rotation of $x_1 y_1 z_1$ by an angle $\alpha$ around the $z$-axis (normal to the plate surface) to get $x_2 y_2 z_2$.

Here we consider a TE wave incidence along the negative $z$-axis direction, with the plane of incidence fixed in the $x_2 z_2$ plane. We extended the enhanced transmittance matrix approach[32] to permit conversion of polarization and applied it to compute the EM fields inside the medium. The EM fields in the crystal can be assumed as the following form

$$\mathbf{E} = \mathbf{S}(z)\exp(j\omega t), \mathbf{S}(z) = (S_x, S_y, 0) \tag{3}$$

$$\mathbf{H} = -j\left(\frac{\varepsilon_0}{\mu_0}\right)^{1/2} \mathbf{U}(z)\exp(j\omega t), \mathbf{U}(z) = (U_x, U_y, 0) \tag{4}$$

where $\varepsilon_0$ and $\mu_0$ are the permittivity and the permeability in vacuum. Substituting (3) and (4) into the Maxwell equations and setting $z' = \omega\sqrt{\varepsilon_0 \mu_0}$, we get the following differential equations

$$\frac{d}{dz'}\begin{pmatrix} S_x \\ S_y \\ U_x \\ U_y \end{pmatrix} = A \begin{pmatrix} S_x \\ S_y \\ U_x \\ U_y \end{pmatrix}, \quad A = \begin{pmatrix} 0 & 0 & 0 & -1 \\ 0 & 0 & 1 & 0 \\ -\varepsilon_{xy} & -\varepsilon_{yy} & 0 & 0 \\ \varepsilon_{xx} & \varepsilon_{xy} & 0 & 0 \end{pmatrix} \quad (5)$$

From the coefficient matrix $A$, we can see that TE wave and TM wave decouple when $\varepsilon_{xy} = 0$, which means conversion between them cannot occur. Only when $\varepsilon_{xy} \neq 0$ should TE and TM waves be solved simultaneously for inclusion of polarization conversion. This proves that $\varepsilon_{xy}$ is the key for polarization conversion. For quantitative description of polarization conversion, we defined the polarization conversion ratio (PCR) as $PCR = T_{sp}/T_{ss} + T_{sp}$, with $T_{ss}$ and $T_{sp}$ the transmittance for the TE wave component unconverted and the TM wave component converted from TE wave, respectively. We calculated the PCR with the uniaxial crystal plate of thickness $d = 0.1$ μm, which is small enough to reduce the influence of wave interference inside the plate. The results are shown in Fig. 2(a) for the wavenumber between 600 cm$^{-1}$ and 1600 cm$^{-1}$ and for two values of the rotation angle $\alpha$. It can be seen that when $\alpha$ is equal to zero, the PCR is completely zero since, in this case, the value of $\varepsilon_{xy}$ is zero. But when $\alpha$ is equal to 45º, the PCR curve shows that conversion of polarization occurs in the plate and two peaks appear at wavenumbers 780 cm$^{-1}$ and 1370 cm$^{-1}$. Furthermore, the value of $|\varepsilon_{xy}|$ as a function of the wavenumber is plotted in Fig. 2(b) for $\alpha = 45º$, which shows that there appear two peaks at the same wavenumbers as in Fig. 2(a). In fact, these two wavenumbers are the resonant wavenumbers of $\varepsilon_{xy}$. Therefore, we can draw a conclusion that the PCR depends on the magnitude of $\varepsilon_{xy}$. Note that the best PCR should be obtained at $\alpha$ equal to 45º, since the absolute value of $\varepsilon_{xy}$ is proportional to $\sin 2\alpha$. In the work by Hao et al.[1], polarization conversion of a linearly polarized wave at normal incidence was studied experimentally after reflection by an anisotropic MM plate. This MM plate has a relative permittivity $\varepsilon=1$ and a relative permeability μ in the form of a diagonal tensor. It was found that a complete polarization conversion between TE and TM waves can be realized when the azimuthal angle of the plane of incidence is equal to 45°. This condition is equivalent to performing a rotational transformation of the coordinates, as discussed above. The rotational transformation also results in an off-diagonal element $\mu_{xy} = 0.5(\mu_{xx} - \mu_{yy})\sin 2\alpha$ of the permeability tensor, which is the key for polarization conversion. The dependence of $\mu_{xy}$ on $\sin 2\alpha$ explains why the complete conversion of polarization occurs at $\alpha = 45^o$. Therefore, our analysis based on rotational transformation can explain clearly the results in Ref.1 and also those in Ref.3-5.

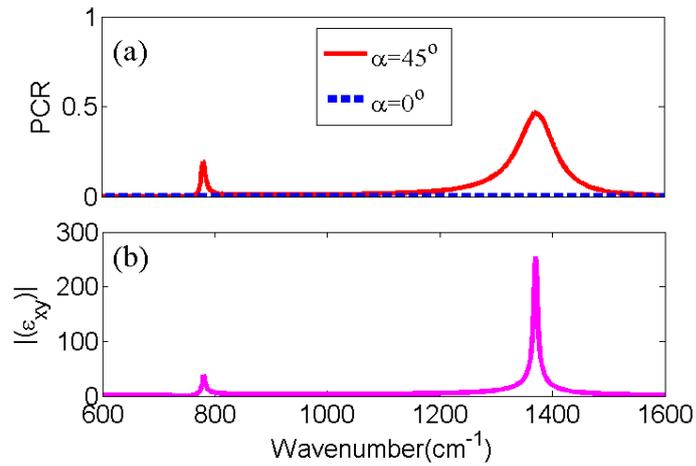

Fig. 2 (Color online) (a) Variation of the PCR with wavenumber for $\alpha$ equal to $0^o$ and $45^o$, (b) Variation of $|\varepsilon_{xy}|$ with wavenumber for $\alpha = 45^o$.

Next we study the impact of the plate thickness $d$ on the PCR. Here, we assume that the wavenumber of the normally incident TE wave is fixed at 1680 cm$^{-1}$. The

PCR varying with the rotation angle $\alpha$ for $d$ in the range from 1 μm to 5 μm is shown in Fig. 3(a). It can be seen that for each value of $d$, the PCR indeed follows closely the relationship of $\sin 2\alpha$ with $\alpha$. Furthermore, the PCR increases with $d$ and a complete conversion is obtained when $d$ is equal to 5 μm and $\alpha = 45^\circ$. However, further increase of $d$ will result in the PCR exhibiting a periodic oscillation with $d$ due to the effect of wave interference inside the plate, as shown in Fig. 3(b) for $\omega = 1680$ cm$^{-1}$ and $\alpha = 45^\circ$. Decrease of the oscillation amplitude with increase of $d$ is due to increase of absorption inside the plate.

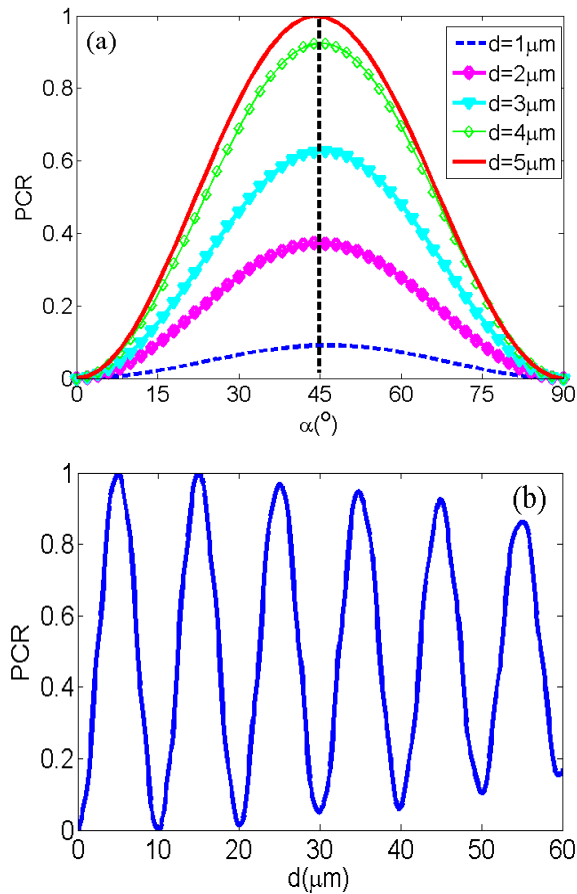

Fig. 3 (Color online) (a) Variation of the PCR with the rotation angle $\alpha$ and thickness $d$ for fixed wavenumber 1680 cm$^{-1}$. (b) variation of the PCR with $d$ for fixed wavenumber 1680 cm$^{-1}$ and $\alpha = 45^\circ$.

From the above analysis we have known that the off-diagonal element $\varepsilon_{xy}$ is the key for conversion of polarization and the thickness of the plate can control effectively the PCR. It should be emphasized here that by selecting carefully the thickness of the plate and the rotation angle, a complete conversion of polarization between TE and TM waves can be achieved in a band. Figure 4 shows the transmittance $T_{sp}$ and $T_{ss}$, and the PCR varying with the wavenumber for $d=3.8$ μm and $\alpha=45^o$. It can be seen that the transmittance $T_{sp}$ is above 0.5 while $T_{ss}$ stays below 0.1 in the wavenumber range from 850 cm$^{-1}$ to 1100 cm$^{-1}$, resulting in the PCR higher than 0.8 in this range. Particularly, the transmittance $T_{ss}$ is almost completely equal to zero in the wavenumber range from 867 cm$^{-1}$ to 933 cm$^{-1}$, such that the PCR is almost completely equal to 1 in this range. This complete conversion is not caused by resonance as the case shown.[1]

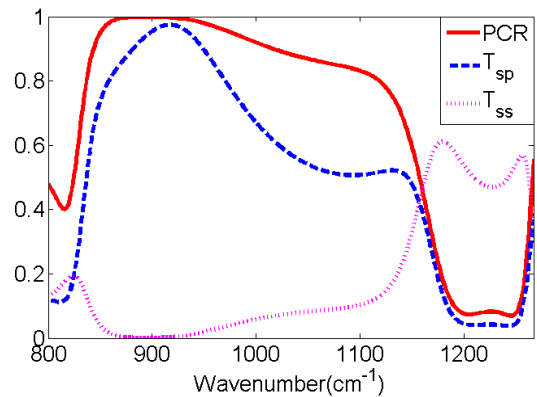

Fig. 4 (Color online) Variation of $T_{ss}$, $T_{sp}$ and the PCR with wavenumber for $d=3.8$ μm and $\alpha=45^o$

Recently, MMs and chiral structures[24-30] have been used to realize optical unidirectional transmission based on polarization conversion. However, these structures are very difficult to fabricate. Here, we propose a simple device that can fulfill unidirectional transmission. The device is constructed by attaching on one side of the uniaxial crystal slab with a one-directional (1D) Ag grating, with the grooves of which perpendicular to the plane of incidence. The period, thickness and filling ratio of the grating are taken as $\Lambda_g$=0.3 μm, $d_g$=0.1 μm and $f$=0.1, respectively. The dielectric function of Ag is described using the Drude model, i.e., $\varepsilon_{Ag} = \varepsilon_\infty - \omega_p^2/(\omega^2 + j\omega\Gamma)$, with $\varepsilon_\infty = 3.4$, $\omega_p = 1.39 \times 10^{16}$ rad/s and $\Gamma = 2.7 \times 10^{13}$ rad/s.[33] The thickness of the uniaxial crystal slab is set as $d$=0.3 μm. We extended the rigorous coupled-wave analysis (RCWA)[32] algorithm to permit conversion of polarization and used it to calculate the transmittance and the absorptance of the device. Assuming a TM plane wave incident on either side of the device, the calculated transmittance and absorptance is shown respectively as a function of the wavenumber in Fig. 5. Clearly, the results reveal that when the wave is incident on the slab side, the absorptance is closed to zero while the transmittance is above 0.8 in the whole specified range of wavenumber. But on the other hand, when the wave is incident on the grating side, the absorptance is greatly enhanced to nearly 0.5 while the transmittance is greatly reduced to below 0.15 at wavenumber 30413.6 cm$^{-1}$. Therefore, quasi unidirectional transmission is realized at this wavenumber, with the difference between the transmittance for incidence from opposite sides of the device up to 0.78. We attribute the physical mechanism for the unidirectional transmission to excitation of surface plasma polaritons (SPPs)[29] on the surface of the Ag grating. When the TM wave is incident on the grating side, SPPs are excited by the diffracted wave at wavenumber 30413.6 cm$^{-1}$ and thus causing large absorption in the grating. As a consequence, the transmittance is reduced. But when the TM wave is incident on the slab side, this TM wave is converted to TE wave when it travels through the slab. We select carefully the thickness of the slab such that the TM wave is almost completely converted into TE wave when it reaches the grating. However, SPPs cannot be excited by TE wave on the surface of the 1D grating. Thus the absorptance in the grating cannot be enhanced and resulting in the high transmittance through the device.

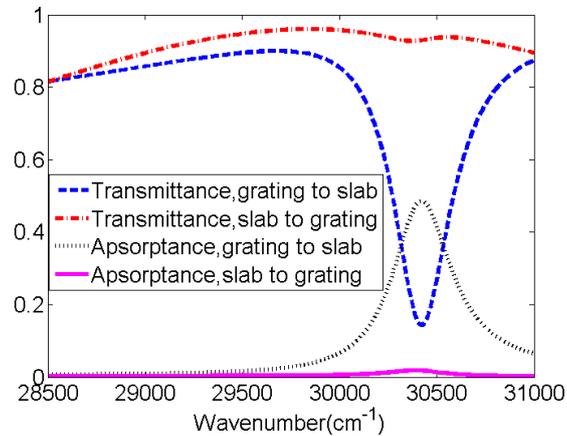

Fig. 5 The spectral-normal transmittance and absorptance of the structure for a TM plane wave incidence from opposite sides of the structure.

In conclusion, we proposed to use a single thin slab of uniaxial crystal to flexibly manipulate the polarization of EM waves by performing a rotational transformation of the coordinates and controlling its thickness. We show that complete conversion of polarization between TE wave and TM wave can be achieved in a spectral band. Finally we propose a simple device composed of a uniaxial crystal slab and a 1D metallic grating to realize optical unidirectional transmission based on polarization conversion and excitation of SPPs.

This work was supported by the National Natural Science Foundation of China (Grant No. 51576004).